\documentstyle[12pt]{article}
\input epsf
\ifx\epsfbox\UnDeFiNeD
\def\figin#1{\vskip2in}
\else\def\figin#1{#1}\fi
\def\tfig#1{{\xdef#1{Fig.\thinspace\the\figno}}
Fig.\thinspace\the\figno \global\advance\figno by1}
\newcommand{\be}{\begin{equation}}
\newcommand{\ee}{\end{equation}}

\newcommand{\ba}{\begin{eqnarray}}
\newcommand{\ea}{\end{eqnarray}}

\newcommand{\r}{\rho}
\newcommand{\s}{\sigma}
\newcommand{\ta}{\tau}
\newcommand{\th}{\theta}

\newcommand{\D}{\Delta}

\newcommand{\p}{\partial}

\newcommand{\CL}{\cal L}

\newcommand{\CP}{\cal P}

\newcommand{\CR}{\cal R}

\newcommand{\Tr}{\rm Tr}

\newcommand{\ra}{\rightarrow}

\def\a{\alpha}

\def\m\mu
\def\n{\nu}
\def\s{\sigma}

\font\ninerm=cmr9

\begin{document}

\hsize36truepc\vsize51truepc
\hoffset=-.4truein\voffset=-0.5truein
\setlength{\textheight}{8.5 in}

\begin{titlepage}
\begin{center}
\hfill LPTENS-99/32\\
\hfill September 7 1999\\

\vskip 0.6 in
{\large  Universality of correlations of levels with discrete statistics}
\vskip .6 in
       {\bf Edouard Br\'ezin \footnote{$ ^\diamond ${\it
brezin@physique.ens.fr}}\\}
            {\it and \\}
       {\bf Vladimir Kazakov  \footnote{$ ^\bullet ${\it
kazakov@physique.ens.fr} }}
\vskip 0.3cm

        \vskip 0.3 cm
{\it Laboratoire de Physique Th\'eorique de l'\'Ecole Normale
Sup\'erieure$^b$\\
24 rue Lhomond, 75231 Paris Cedex 05, France\\}

\vskip 0.6 in

{\bf Abstract}
\end{center}
\vskip 14.5pt

{\leftskip=0.5truein\rightskip=0.5truein\noindent
{\small


We study the statistics of a system of N random levels with integer values, 
in the presence of a  logarithmic repulsive potential of Dyson type. 
This probleme arises in sums over  representations  (Young tableaux) of  
GL(N) in various matrix problems and in the study of statistics of 
partitions for the  permutation group. The model is generalized to 
include an external source and its correlators are found 
in closed form for any N.
We reproduce  the density  of levels in the large N and  double scaling 
limits and the universal correlation functions  in Dyson's short-distance 
scaling limit. We also study the statistics of small levels.  

}
\par}

\vskip 1.8in
\hrule width5cm \vskip 2pt
{\ninerm \hskip -17pt
$^b$ Unit\'e Mixte de Recherche 8549 du Centre National de la Recherche
Scientifique et de l'\'Ecole Normale Sup\'erieure}

\newpage

\end{titlepage}
\setlength{\baselineskip}{1.5\baselineskip}


\section{  Introduction and definition of the generalized model}

The theory of random matrices leads often to consider character expansions,
i.e. sums over irreducible representations of groups
such as $GL(N)$ or $U(N)$
\cite{MIG,RUS,DUKA,WYNK,GRMA,KSW,STWKO,DIFITZ,KAZI}.
Similar sums play also a role
in probability theory  \cite{VER}, in particular
in recent studies of  the distribution of cycles in the group of
permutations \cite{DEIFT,BOO,BOD,JOHO,JOHD}. This
has led to the study of probability distribution functions (PDF) for  $N$
non-negative integers $h_1,\cdots,h_N$
of the form :
\be
{\CP}_{\a}(h_1,\cdots, h_N) \sim
  \D^2(h) \a^{\sum_{k=1}^N h_k}
\label{PDF}
\ee
where $\D(h)=\prod_{m>j}(h_m-h_j)$, and $0<\a<1$ is a real parameter.

Let us generalize this distribution to an ensemble characterized by $N$
parameters
$\a_1,\cdots,\a_N$, in the following way :
\be
{\CP}_{[\a_1,\cdots,\a_N]} (h_1,\cdots, h_N) =
\frac{1}{Z} \D(h) \chi_h(\{\alpha\})
\label{PDFG}
\ee
where $\displaystyle\chi_h(\alpha)={\det_{k,j}\a_k^{h_j}\over
\D(\a)}$ is the Weyl character of a
diagonal $GL(N)$ group element $\a_1,\cdots,\a_N$ of a given
irreducible representation $R$. $R$ is fixed in terms of the highest
weights $m_k=h_k+k-N, \ \ k=1,\cdots,N$. (In principle the corresponding
$h_k$'s are strictly ordered,
but in view of the symmetry of the weight (\ref{PDF}) this restriction may
be ignored in the sums).
 The constant $Z$ is defined by the normalization condition
\be Z = \sum_{h_1=0}^\infty \cdots\sum_{h_N=0}^\infty \Delta(h)
 \chi_h(\{\alpha\})
\ee

Note that in the limit of
coinciding $\a$'s $\a_k\ra\alpha$ the (non-normalized) distribution
(\ref{PDFG}) reduces to (\ref{PDF});
this follows from the well known formula:
\be
  \chi_h(\alpha) \ra \prod_{k=1}^N \a^{\sum_k m_k} \dim_{\{m\}}=
 { \Delta(h)\over \prod_{k=0}^{N-1}k!}\a^{\sum_k (h_k-k+N)}
\label{LICH}
\ee
where $\dim_{\{m\}}$ is the dimension of a representation given by the
highest weights $m_1,\cdots,m_N$.

There are several reasons for this generalization. First it leads to simple
exact formulae, as in the case of random
matrices coupled to an external matrix source \cite{KAZ,BH}; the $N$
parameters $\a_k$'s play here
the role of the eigenvalues of the source. Here also, even when the source
is a simple multiple of the identity, meaning now that all the $\a_k$'s
are equal to a single $\a$, the final formulae are explicit and simple.
Furthermore  these parameters $\a_k$'s provide a powerful check of
universality. Indeed
it is  found here that at generic points, at which the density of levels is
non-singular, the correlations are, in the proper Dyson scaling limit,
insensitive
to the specific probability distribution. Singular points fall also into
universal classes, as for the usual matrix models \cite{KAZ}.
Varying the external
parameters $\a_k$   one can indeed tune various singular classes : examples
are given in the subsequent sections.
The machinery developed for integrable systems, and used in
\cite{DEIFT,BOO,BOD,JOHO,JOHD} for the single- parameter model (\ref{PDF}),
ought to be be
applicable to our generalized model (\ref{PDFG}) as well.

Finally  the
distribution probability (\ref{PDF}) provides a natural measure on the
$S_\infty$ group of arbitrary permutations ; the $m_k$'s with $\
k=1,2,\cdots,N$, are the lengths of the cycles of a
permutation class consisting exactly of $N$ cycles.
One can also interpret this distribution probability as defined on the
(infinite) set of all Young tableaux with $m_k$ boxes in the $k$'th
row.

In our case, (\ref{PDFG}) is a natural
multi-parametric generalization of (\ref{PDF}), which may now be interpreted
as a specific coloring of Young tableaux.
The boxes of the
Young tableau characterizing a permutation, are colored in  $N$ colors in a
way which is explained  below ; the $k$-th  color is weighted with $\a_k$.
For instance,
for $N=2$,
there are two rows of lengths (highest weights) $m_1\ge m_2$ in the
corresponding Young tableau characterizing a class of permutations
with 2 cycles,
and for the character we have the following finite sum of positive
terms:
\ba
\chi_{m_1,m_2}(\a_1,\a_2) &=
\sum_{k=0}^{m_1-m_2}\a_1^{m_1-k}\a_2^{m_2+k}= \cr
&= \a_1^{m_1}\a_2^{m_2}+\a_1^{m_1-1}\a_2^{m_2+1}+\cdots
+\a_1^{m_2+1}\a_2^{m_1-1}+\a_1^{m_2}\a_2^{m_1}
\label{CHEX}
\ea
This may be interpreted as a coloring of a tableau, in which   the first
$m_1-k$ boxes of the upper row have a color 1 and
 all the other boxes have color 2 ; we sum  over $k$ with a
factor $\a_1^{\#1}\a_2^{\#2}$ where $\#1,\#2$ are the numbers of boxes
(``areas'') of the Young tableau of a given color.

To generalize this formula to any $N$, we have to expand a general character
into a sum of  monomials in the $\a$'s (generators of the maximal torus of
$GL(N)$):
\be
\chi_{\{m\}}=\sum_{\{l\}_{\{m\}}} n\{l\} \prod_{k=1}^N(\a_k)^{l_k}
\label{DEFCH}
\ee
where the finite sum over the positive integers
$(l_1,\cdots,l_N)$ characterizing the elements of the Lie algebra of
the maximal torus, or the weights of representation R, is restricted in
a specific way by the shape of the corresponding Young tableau of the
representation $R(m_1,\cdots,m_N)$. The positive integers $n\{l\}$ are
called the multiplicities of those weights.

To make all this more explicit, and to give it a nice probabilistic
interpretation, let us introduce the Gelfand-Tseytlin scheme (GT
scheme), which provides  an orthonormal basis of states for a given
representation
$R$ of $GL(N)$ characterized by the highest weights $m_i, \
i=1,\cdots,N$. Every state vector of $R$ is characterized by
$N(N-1)/2$ positive integers $m_{kj}, k=1,\cdots,N, \ j=1,\cdots,k$. The
first ones,
 the $m_{N,i}\equiv m_i, \ i=1,\cdots,N$,  are simply the highest weights.
The subsequent ones are  given integers, restricted by
 the inequalities:
\be
 m_{j,i}\ge m_{j-1,i}\ge m_{j,i+1}, \ \ i=1,\cdots,j
\label{INEQ}
\ee
The basis vectors of this representation are thus characterized by a
triangular array usually depicted as \cite{BARA}
\ba
m_{N,1}&\ \hspace {8mm} m_{N,2} \hspace {8mm}  \hspace {8mm}\cdots
\hspace {12mm}m_{N,N}\nonumber \\
& \  m_{N-1,1} \hspace {8mm} \cdots \hspace {8mm}m_{N-1,N-1} \nonumber  \\
&\vdots \hspace {20mm}\vdots\nonumber \\ &\ m_{N,N}
\ea
in which any $m$ in a given row lies in-between  the two $m$'s next to it
in the previous row.
We will use the definition of a character as a trace of a group
element  in its diagonal form, for a given representation $R$:
$\chi_R(\{\a\})=\Tr_R \prod_{j=1}^N\a_j^{T^R_{jj}}$, where $T^R_{jj}$
is a diagonal generator in the Lie algebra of  this representation.
In the GT basis the values of $T^R_{jj}$ are given as (see for example
\cite{BARA})
\be
T^R_{jj}\equiv l_i=\sum_{i=1}^jm_{ji}-\sum_{i=1}^{j-1}m_{j-1,i}
\label{LDEF}
\ee
The  expansion (\ref{DEFCH}) is now explicitely given by
the formula:
\be
\chi_R(\{\a\})=\sum_{\{GT\}_R} \prod_{j=1}^N \a_j^{l_j\{m\}}
\label{CHSOM}
\ee
where the sum is taken over all GT schemes (states), satisfying  the
restrictions (\ref{INEQ}). Of course in this sum the same monomial may
appear several times. The number of times each monomial enters
in the sum is  the multiplicity $n\{l\}$ of eq. (\ref{DEFCH}).

This formula may be given an interesting probabilistic geometrical
interpretation, in terms of sums over the colorings of Young tableaux.
Let us define a coloring in the following way (see figure 1 for an
illustration of the $N=3$ case).  The color number one is given to the
last $m_{N,1}-m_{N-1,1}$ boxes of the first row, the last
$m_{N,1}-m_{N-1,1}$ boxes of the second row, etc., the $m_{N,N}$ boxes
of the last row. Color number 2 is given to the boxes numbered
$m_{N-1,1}+1$ to $m_{N-2,1}$ in the first row, $m_{N-1,2}+1$ to
$m_{N-2,2}$ in the second row, etc., the first $m_{N-1,N-1}$ boxes of
the row before last ; etc., the $N$-th color is given to the first
$m_{1,1}$ boxes of the first row.  At the end, the Young tableau is a
"brick wall" made out of colored bricks, each brick at an upper level
relying on zero, one or two lower bricks of lower color indices, as
depicted on fig.1.
%
\vskip 50pt
\hskip 20pt
\epsfbox{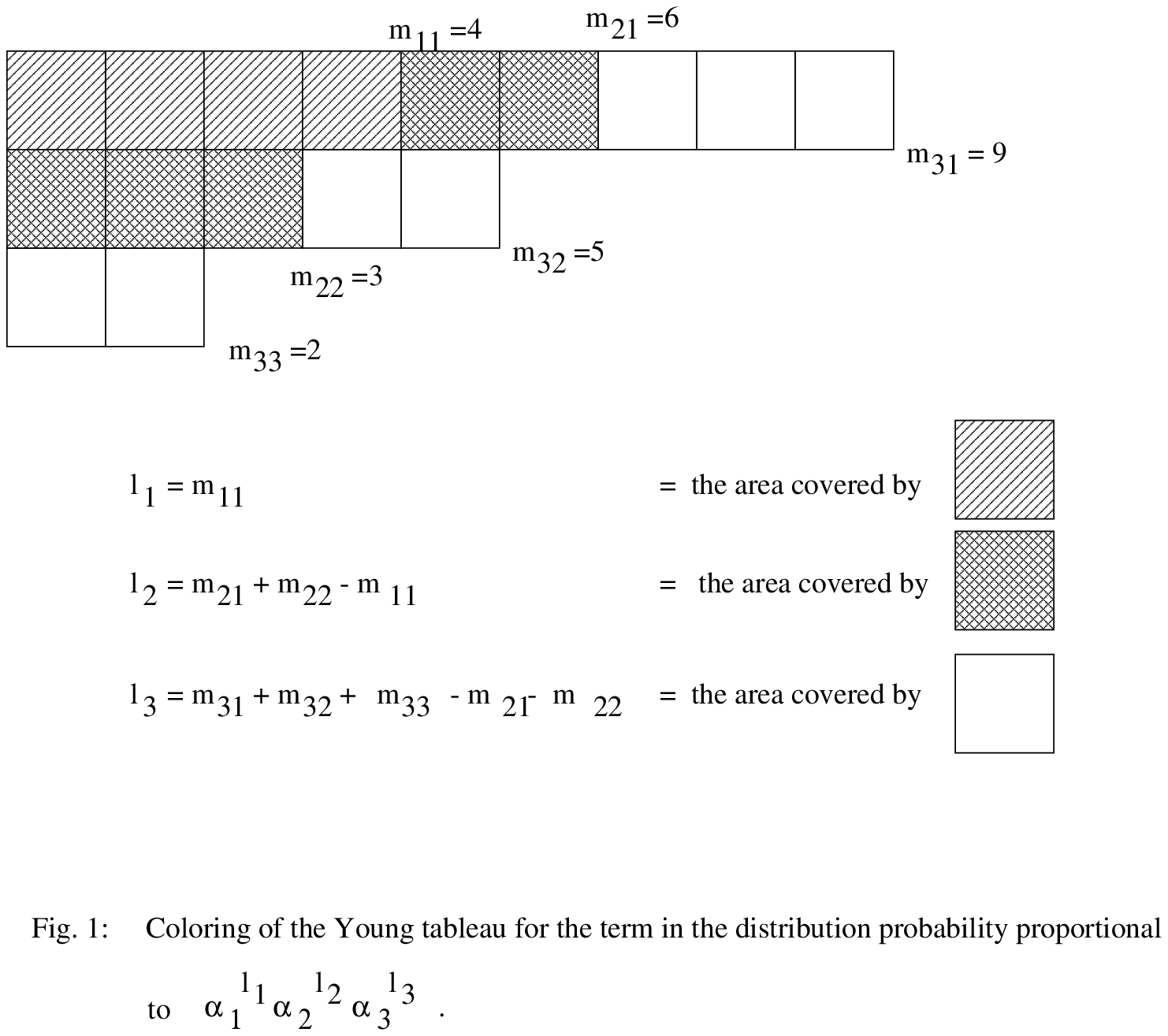}
\vskip 20pt

\bigskip

Note that if all the  $\a_k$'s reduce to $\a_k=1$, one is simply summing
in (\ref{CHSOM}) over each state  with weight one ; therefore one
recovers the dimension of the representation $dim_R$ as stated
in (\ref{LICH}).

If one substitutes the expression (\ref{CHSOM}) for the character into
(\ref{PDFG}), one  obtains a  probability for the colored Young tableaux
(related to  permutations of colored objects).

In the next sections, we derive explicit (for any N)
formulae for the correlators of the distribution (\ref{PDFG}) and study
various scaling limits.

\section{   Density of discrete levels}

The density of levels is defined as
\be
 \rho(\sigma)=\langle {1\over N}\sum_k\delta_{N\sigma,h_k}\rangle =
\oint{dt\over 2i\pi
t^{N\sigma+1}} \langle {1\over N} \sum_k t^{h_k}\rangle \ ,
\label{DEN}
\ee
in which the brackets denote the averaging with respect to the
probability distribution (\ref{PDFG}). The variable $N\sigma$ is a priori
an integer but, in the large
$N$-limit,
$\sigma$ will be considered as a continuous variable.
Let us calculate $ U_1(t)\equiv \langle {1\over N} \sum_k
t^{h_k}\rangle_h$  by performing
explicitly the sums over the $h$'s. Using the antisymmetry of $\Delta(h)$
in (\ref{PDFG}) we can rewrite it as
\be
 U_1(t)
=C \sum_k \sum_{h}t^{h_k}{\det_{m,j}\a_m^{h_j} \over
\Delta(\a)}=
{C\over\Delta(\a)}\sum_k \sum_P(-)^P
\sum_{h}\prod_m\left(\a^{(k)}_m\right)^{h_m} h_m^{P_m-1}
\label{DENT}
\ee
where the determinant is represented as a sum over permutations $P$ of
the levels $h_1,\cdots,h_N$ and by definition
\be
\a^{(k)}_m=t^{\delta_{m,k}}\a_m .\ee
The sums over $h$'s are now independent and can be calculated using
the formula :
$$\sum_{h=0}^\infty h^p\a^h=(\a \p_\a)^p {1\over 1-\a}=
{1\over p!}\ {1\over 1-\a}Q_p\left({1\over 1-\a}\right)  $$
where $Q_p(x)$ is a polynomial of degree $p$, whose coefficient of highest
degree is one. Since for any set of such
polynomials  $\det_{p,q} Q_{p-1}(x_q)=\Delta(x_1,\cdots,x_N)$ we obtain
from (\ref{DENT}) :
\be
  U_1(t)=
{C\over\Delta(\a)}\sum_k {\Delta\left(
(1-\a^{(k)})^{-1}\right) \over \prod_m \left(
1-\a^{(k)}_m\right)}
=  {1\over N}\sum_k \left[{1-\a_k\over
1-t\a_k}\right]^N \prod_j'  \ {t\a_k-\a_j\over
\a_k-\a_j}
\label{DENC}
\ee
where prime in the product means that the term  $k=j$  is omitted.
The constant $C$ has been determined, through the normalization condition
$U_1(1)=1$.


This quantity has an elegant contour integral representation, similar to
the one found in \cite{KAZ} for  integrals over hermitian random matrices
with an external matrix source. Indeed, the integral
\be
  U_1(t) =
{1\over N(t-1)} \oint {du\over 2i\pi u}
\left({1-u\over 1-tu}\right)^N
 \prod_j{t u-\a_j\over
u-\a_j},
\label{INR}
\ee
over a contour in the complex u-plane which surrounds the N poles
$\alpha_j$'s (and not the origine), reproduces
the sum (\ref{DENC}).
For the density of levels (\ref{DEN}), since
\be
\rho(\sigma) = \oint \frac{dt}{2i\pi} \frac {U_1(t)}{t^{N\sigma+1}},
\ee
where the integration contour encircles the vicinity of the origin,
we obtain, using (\ref{INR}):
\be
 \rho(\sigma)=  {1\over N} \oint {dt\over 2i\pi t^{N\sigma+1}(t-1)}
\oint {du\over 2i\pi u}\left({1-u \over 1-tu}\right)^N
\prod_j {t u-\a_j\over u-\a_j}.
\label{DENI}
\ee
Similarly for the resolvent, defined as $\displaystyle G(z)=\langle{1\over
N}\sum_{k=1}^N{1\over z-h_k/N }\rangle$,  we can use
\be G(z) = \int_0^{\infty} d\tau e^{-\tau z}\  U_1(e^{ \tau/N})\ee
and obtain the integral
representation:
\be
 G(z)=  {1\over N} \int_0^\infty {d\ta\over (e^{\ta/N}-1)}e^{-\ta z}
\oint {du\over 2i\pi u}\left({1-u \over 1-e^{\ta/N} u}\right)^N
\prod_j {e^{\ta/N} u-\a_j\over u-\a_j}
\label{CIG}
\ee

After the change $t \ra t/u$ in (\ref{DENI}) we get ($\s=p/N$):
\be
\rho(p/N)=  {1\over N} \oint {dt\over 2i\pi t^{p+1}}\prod_j {(t-\a_j)
\over (t-1)}
\oint {du u^{p+1}\over 2i\pi }\prod_j {(u-1)\over  (u-\a_j)}
{1\over t-u}
\label{DENTW}
\ee
or, inflating the contour of integration in $u$ and changing $u\ra
1/w$,:
\be
\rho(p/N)=  {1\over N} \oint {dt\over 2\pi t^{p+1}}\prod_j {(\a_j-t)
\over (1-t) }
\oint {dw \over 2\pi w^{p+2}}\prod_j {(1-w)\over  (1-\a_j w)}
{1\over tw-1}
\label{DENTI}
\ee

Expanding the last factor we get a finite sum representation for the
correlator:
\be \rho(p) =\frac{-1}{N}\sum_{k=0}^p
{\CL}^{p-k}_N(\a)\hat{\CL}^{p-k+1}_N(\a)
\label{PRO}
\ee where the polynomials in $\a$'s ${\CL}^{q}_N(\a)$ and
$\hat{\CL}^{q}_N(\a)$ are defined as
\be
{\CL}^{q}_N(\{\a\})=\oint {dx
\over 2\pi i x^{q+1}} \prod_j{(\a_j-x)\over (1-x)}
\label{POLN}
\ee
\be
\hat{\CL}^{q}_N(\{\a\})=\oint {dx \over 2\pi i x^{q+1}}   \prod_j{(1-x)\over
(1-\a_j x)}
\label{POLNH}
\ee
The contours for both (\ref{POLN}, \ref{POLNH}) encircle the pole at
$x=0$.

Note that this expression is exact, for any $N$ and any set of $\a$'s. In
particular  for the density, in the simplest case
$\a_1=\a_2=\cdots=\a_N=\a$,   most studied in the literature, this leads to
a simple integral:
\be
\rho(p/N)=  {-1\over N} \oint {dt\over 2\pi t^{p+1}}{(t-\a)^N
\over (t-1)^N}
\oint {du u^{p+1}\over 2\pi }{(u-1)^N\over  (u-\a)^N}
{1\over t-u}
\label{DENCR}
\ee
or, after inflating the contour and changing $u\ra \a/w$
\be
 \rho(p/N)= -\oint  {du\over 2i\pi u}
\oint {dt\over 2i\pi t^{N\s+1}(t-1)}
\left({(1-u)(\a-tu)\over (1-tu)(u-\a)}\right)^N.
\label{DENL}
\ee

The polynomial (\ref{POLN}) reduces now to:
\be
{\CL}^{q}_N(\a)=\oint {dx \over
2\pi i x^{q+1}} \left({\a-x\over 1-x}\right)^N
\label{POLEQ}
\ee
and the two sets of polynomials are related here simply by
\be
{\CL}^{q}_N(\a) = \a^{N-q}\hat{\CL}^{q}_N(\a).
\ee

In the next sections we will study the large $N$ limit of this density
and, for the case of equal $\a$'s, compare it with a direct computation
based on the solution of a Riemann-Hilbert problem given in the
appendix.

\section{ Pair correlator of discrete levels  }

The simplest interesting correlator, which is conjectured to obey a short
distance
universality for  nearby levels is the pair correlator:
\be
 U_2(t_1,t_2) \equiv {1\over N^2}
\langle \ \sum_{k\neq l} t_1^{h_k} t_2^{h_l}\ \rangle
\label{DEFZ}
\ee
or rather its connected part:
\be
K_2(t_1,t_2) \equiv  U_2(t_1,t_2)- U_1(t_1) U_1(t_2)
\label{DEFK}
\ee
A similar calculation yields
\ba  U_2(t_1,t_2) =&&
\frac{1}{N^2}\sum_{k,l}\frac{(t_1\a_k-t_2\a_l)(\a_k-\a_l)}{(t_1\a_k-\a_l)(\a_k-t
_2\a_l)}\nonumber\\
&&\prod_{m\neq k} \frac{(\a_m-t_1\a_k)}{(\a_m-\a_k)}\prod_{m\neq l}
\frac{(\a_m-t_2\a_l)}{(\a_m-\a_l)}\nonumber\\&& \prod_k
\left(\frac{1-\a_k}{1-\a_kt_1}\right)^N
\left(\frac{1-\a_k}{1-\a_kt_2}\right)^N,\ea
from which one derives again a contour integral representation over two
complex variables (similar to  the representation for the hermitian
matrix model found in \cite{BH}):
\ba
 K_2(t_1,t_2)
=&&  -{1\over N^2} \oint\oint {du dv\over (2i\pi)^2
(u-t_2 v)(v-t_1 u) }
\left({1-u\over 1-t_1u}\right)^N\left({1-v\over 1-t_2v}\right)^N\nonumber
\\ &&
 \prod_j {(t_1 u-\a_j)(t_2 v-\a_j)\over (u-\a_j)(v-\a_j) }
\label{CIK}
\ea
Finally, for the density correlator
\be
\rho(p,q)=\langle\sum_k\delta_{p,h_k}\sum_m\delta_{q,h_m}\rangle
= \oint \oint{dt_1\over 2i\pi\  t_1^{p+1}}{dt_2\over 2i\pi\ t_2^{q+1}}
K_2(t_1,t_2)
\label{DEFKK}
\ee
we obtain ( after changing variables from $t_1$ to $t_1/u$ and $t_2$ to
$t_2/v$), a factorized formula,
again in similarity
with the hermitian matrix model, \cite{BH}:
\be
 \rho(p,q)=-R(p,q)R(q,p)
\label{FACT}
\ee
where
\be
 R(p,q) =\oint\oint {du dt t^{-p}  u^{q-p-1}\over (2\pi)^2(1-t)}
 \left({u-1\over t-1}\right)^N\prod_j{(t-\a_j)\over (u -\a_j)}
\label{CIRR}
\ee
or, going back to the previous non-scaled variables:
\be
 R(p,q) =-\oint\oint {du dt t^{-p}  u^{q}\over (2\pi)^2(u-t)}
\prod_j{(t-\a_j)\over (t-1)}{(u-1)\over (u -\a_j)}
\label{CIR}
\ee

Another useful form of $R(p,q)$ can be obtained (similarly to
(\ref{DENTI})) by  the contour of integration in $u$
in (\ref{CIR}) and changing $u\ra 1/w$ to catch the poles at $w=0$:
\be
 R(p,q) =\oint\oint {dw dt t^{-p}  u^{-q}\over (2\pi)^2}
 \prod_j{(\a_j-t)\over (1-t)}{(1-w)\over (1-\a_j w)}{1\over (1-tw)}
\label{CIRRI}
\ee

Expanding the last factor we get a finite sum representation for the
correlator:
\be R(p,q) = \a^{-N+q}\sum_{k=0}^{{\rm inf}(p,q)}
{\CL}^{p-k}_N(\a)\hat{\CL}^{q-k}_N(\a)
\label{PROB}
\ee
where ${\CL}^{q}_N(\a)$ and $\hat{\CL}^{q}_N(\a)$ are given by
(\ref{POLN}) and (\ref{POLNH}).

In the next sections we will study these formulae in the large $N$
limit, for the case of large Young tableaux.

\section{Study of the density in the large N limit}

In the large N limit the formula for the resolvent (\ref{CIG}) gives
\be
 G(\s)= \int_0^\infty {d\ta\over \ta}\oint {du\over 2\pi u}
\exp\left(-\ta[\s-{u\over 1-u}-uG_0(u)]\right)
\label{IGN}
\ee
where $G_0(u)={1\over N}\sum_{j=1}^N {1\over u-\a_j}$ is the resolved
of distribution of parameters $\a_j$ of our problem.
Integration over $\tau$ leads to the eq.:
\be
 G(\s)= \oint {du\over 2\pi u}
\log\left([\s-{u\over 1-u}-uG_0(u)]\right)
\label{IGN2}
\ee

Differentiating the last eq. in $\s$, taking into account the
contribution of the pole at
\be
\s={u\over 1-u}+uG_0(u)
\label{NEWV}
\ee
and integrating back in $\s$ we obtain an explicit equation for the resolvent
\be
G(s)=\ln u(s)
\label{EXPG}
\ee
where $u(s)$ is a solution of the eq.(\ref{NEWV}). The constant of
integration in (\ref{EXPG}) is chosen to be zero in order to match the
assymptotics $\s \ra \infty$ ($u\ra 1$).

If we eliminate $u$ from the eqs. (\ref{NEWV}) and  (\ref{EXPG}) we obtain
the following functional equation for $G(\s)$
\be
e^{G(\s)}=1+{1\over\s} +{1\over\s}G_0(e^{G(\s)})
\label{FEG}
\ee
This functional formula is very similar to those found by the direct
analysis of the saddle point equations in
\cite{WYNK,GRMA} for the heat kernel on the group $SU(N)$ in the large
$N$ limit (see also \cite{KSW,STWKO} for many other similar
results).

For the particular case $\a_1=\a_2=\cdots=\a_N=\a$ we have $G_0(u)={1\over
u-\a}$ and $G(\s)$ is given by
\be
G(\s)= \log\left( {1\over 2\s}\left((\a+1)\s-(1-\a)-
(1-\a)\sqrt{[\s-{1-\sqrt{\a}\over 1+\sqrt{\a}}]
[\s-{1+\sqrt{\a}\over 1-\sqrt{\a}}]}\right)\right)
\label{CONL}
\ee
Since  on the real axis
\be
G(\s\pm 0)={\it Re} G(\s)\pm i\pi\rho(\s)
\label{GRA}
\ee
 we see
from (\ref{CONL}) that
\be
\rho(\s)=1, \ \ {\it for} \ \ 0<\s< b
\label{SAT}
\ee
\be
\rho(\s)=\log\left( {1\over 2\s}\left((\a+1)\s-(1-\a)-
(1-\a)\sqrt{(\s-b)(a-\s)}\right)\right),
{\it for} \ \ b<\s<a
\label{COND}
\ee
where
\be
a={ 1+\sqrt{\a}\over 1-\sqrt{\a}}, \ \ \ \ \ b= {1-\sqrt{\a}\over
1+\sqrt{\a}}.
\label{LIM}
\ee
This coincides with the direct saddle point calculation of the Appendix
and reproduces an old result of A. Vershik and S. Kerov \cite{VER}
on the limiting  shape of the Young tableau.

Another interesting problem corresponds to the statistics of the weights
close to the upper or lower end of the distribution
\cite{BOO,BOD} etc. studied in the next section.

\section{ Double scaling limits}

The density function contains three special points ("end points"):
$z_c=0,a,b$,  near which it exhibits a universal
behavior.  We will study the close vicinity of the points $a,b$ such
that the deviation $\Delta z$ from one of these points scales with $N$
according to the power law $\Delta z \sim N^{-2/3}$ typical of the
double scaling limit for the generic distributions of the eigenvalues
of matrix models \cite{BRKA}.
The scaling for $\a \ra 1$ in the vicinity $z\sim b \sim 0$ will
be different and will be considered afterwards.

It is useful to put in (\ref {CIG})
$t=e^{\ta/N}=1+\th/N$. This leads to the integral
representation:
\be
 G(z)=
{1\over N} \int_0^\infty {d\th\over \th(1+\th/N)^{zN+1}}
\oint {du\over 2\pi u}\left(1-{u \over 1-u}{\th\over N}\right)^{-N}
\prod_j \left(1+ {u\over u-\a_j}{\th\over N}\right)
\label{CIGTH}
\ee

Exponentiating the integrand and expanding in $1/N$ we get:
\ba
G(z)=
\int_0^\infty {d\th\over {\th(1+\th/N)}}\oint {du\over 2\pi u}\exp
\big[-\th\left( z -{u\over 1-u}-uG_0(u)\right)     \cr
+{\th^2\over 2N}\left(z
+{u^2\over (1-u)^2}+u^2G'_0(u)\right)-
+{\th^3\over 3N^2}\left( -z +{u^3\over
(1-u)^3}+\frac{1}{2}u^3G''_0(u)\right)+ \cdots \big]
\label{GEXP}
\ea

The end points $a,b$ are defined by the equation (\ref{NEWV}).  The
singular behavior arises when the linear in $\th$ term in (\ref{GEXP})
develops a double zero at the the end point (i.e., when its
derivative in $u$ is zero):
\be
G_0(u)+uG'_0(u)+{1\over
(1-u)^2}=0
\label{DOZ}
\ee

One sees immediately that if the conditions (\ref{NEWV}) and
(\ref{DOZ}) are satisfied the quadratic in $\th$ term is also
zero.In the vicinity of the end points  $z=a$ or $z=b$ we can expand
in the exponential  in powers of $\Delta=z-a$ (or respectively $(z-b)$) and
$\delta u=u-u_c$:
\ba
 \frac{\partial}{\partial z}G(z)= \cr
- \int_0^\infty {d\th}\oint
{du\over 2i\pi u}\exp \big[-\th \left(\Delta
-\frac{1}{2}A(\delta u)^2\right)
+{\th^2\over 2N}\left(\Delta +A u_c \delta u\right) +Au_c^2 {\th^3 \over
6N^2}+\cdots\big]
\label{CRID}
\ea
with $\displaystyle A= -\frac{2z_c}{u_c^2} +\frac{2u_c}{(1-u_c)^3} +u_cG_0''$.
In the double scaling regime, when $\Delta\sim N^{-2/3}$, the
quadratic term in $\th^2\Delta/N$ terms is negligible, and after Gaussian
integration in $u$ along the contour of the stationary phase, we
obtain an Airy-like function of $\D N^{2/3}$
\be
\frac{\partial}{\partial z}G(z)\simeq \int_0^\infty {d\th\over \th^{1/2}}
e^{-\D N^{2/3}\th-B{\th^3}}.
\label{DSL}
\ee

This gives for the density a  double scaling expression:
\be
\rho(\s) \sim \Delta^{1/2} f(\Delta N^{2/3})
\ee
where the function $f(x)$ of the double scaling parameter
can be defined nonperturbatively by an appropriate change of the
integration contour  in (\ref{DSL}).

For the last end point of the distribution, namely  the vicinity of  $z=0$,
in the simple
large $N$ limit $\rho(z)=1$ around this point. But it will be not so
any more in the special double scaling limit in which $\a$ approaches 1 and
the interval $(0,b)$,
on which $\rho$ remains fixed to one, shrinks to zero.
We will consider this special limit in section 7.

\section{ Study of  the density correlator for nearby levels }

Now we shall study the quantity $R(p,q)$ in the large $N$ limit for a
separation   of $p,q$ of order one : $|p-q|=1,2,3,\cdots$.
The  result is expected to be universal (if we express $p-q$ in terms of
the local level spacing  $\rho(p)^{-1}$).

 The study of the large $N$ limit of $R(p,q)$ is very similar to the
study of the density $\rho(p)$ in the previous section. Namely we put again
$t=e^{\ta/N}$ and retain in (\ref{CIR}) only the terms of the order
$\sim 1$. The integration over $\ta$ yields the following large N
limit of $R$:
\be
R(p,q)={[u_+(q/N)]^{|p-q|}-[u_-(q/N)]^{|p-q|}\over |p-q|}
\label{RPQ}
\ee
where $u_\pm(\s)=e^{G(\s)}$ are   two conjugated solutions of (\ref{NEWV})
corresponding to two choices of the sign in (\ref{GRA}) when approaching
the real $\s$-axis. Hence we get on the real axis:
\be
R(p,q)=2i e^{2{\it Re} G(q/N)}{\sin\left(\pi\rho(q/N)|p-q|\right)\over
|p-q|}
\label{UNR}
\ee

The level correlation function (\ref{FACT}) is thus
\be
\rho(p,q) \sim {\sin^2\left(\pi\rho(q/N)|p-q|\right)\over
|p-q|^2}
\label{UNIV}
\ee
It reproduces the well-known universal formula for the  level
correlation function with Dyson repulsion law (for the correlations of
eigenvalues of the unitary ensemble of matrices).
It does not change even on the saturated
part of the corresponding Young tableau where $\rho(q/N)=1$.
The only difference with respect to the hermitian ensemble is that
 in our case of discrete levels is that the this correlation
function makes sense only at discrete values of the distance between
levels $|p-q|=1,2,3,\cdots$.

The authors of \cite{BOO,BOD,JOHO,JOHD} came to the same conclusion
in a particular case of equal $\a$'s.
Our generalization to any collection of $\a_k$'s shows once again
a remarkable universality of the classical result (\ref{UNIV}).

\section{A special large N limit: small weights in very large Young tableaux}

In this section we study a new singular scaling regime, in which all the
$\a_k\ra 1$ and $N \ra
\infty$,  so that the parameter $\displaystyle \rho=N(1-{1\over
N}\sum_k\a_k)$ remains finite.

In this limit the polynomial ${\CL}_N^{(p)}(\a)$ defined by (\ref{POLN})
becomes :
 \be
{\CL}_N^{(p)}(\a) \ra \oint {dt\over 2\pi i t^{p+1}}
e^{\rho/(t-1)}=e^{-\rho}\oint {dz
(1+z)^{p-1}\over 2\pi i z^{p+1}} e^{-\rho z}= e^{-\rho}M_p(\r) ,
\label{LAG}\ee
with
\be M_p(\r) = L_p(\r)-L_{p-1}(\r)\ee
in which $L_n(\rho)$ is the standard Laguerre polynomial of order $n$,
normalized to
$L_n(0)=1$.
Similarly, in this limit,
\be \hat{\CL}^{(p)}_N(\a)\ra M_p(\r)\ee
The formula (\ref{PRO}) becomes:
\be
{\CP}_\infty(p) =
{e^{-\rho} }\sum_{k=0}^p M_{k}(\rho) M_{k+1}(\rho)
\label{PROBE}.
\ee
In the limit $\rho\ra 0$ of ultra large Young tableaux, we have
$\displaystyle M_p(\rho)\simeq -\rho$ and the formula (\ref{PROBE})
gives explicitly
\be {\CP}_\infty(p) = \r -(p+1)\r^2 +O(\r^3)\ee

In full analogy with these formulae for the density we can deduce from
(\ref{PROB}) the correlation function of the small weights $p$ and $q$
for large Young tableaux:
\be
{\CR}_\infty(p,q) =
 \sum_{k=0}^{{\rm{inf}}(p,q)}M_{p-k}(\r)M_{q-k}(\r)
\label{CRL}
\ee
which gives for the limit $\r\ra 0$ of ultra large Young tableaux
\be
{\CR}_\infty(p,q)= -\rho + {\rm{inf}}(p,q) \rho^2+ O(\r^3)
\label{RCOR}
\ee

\section {Conclusion}

In this work we have studied  discrete statistics of the kind (\ref{PDFG}).
Such character expansions (CE)  appear in many
situations in which one sums over the
irreducible representations of $GL(N)$ : the hermitian matrix models,
circular ensembles, in various partition functions
and loop averages of gauge theories ,
such as those encountered in two-dimensional quantum chromodynamics
\cite{MIG,RUS}.  The systematic large N saddle point analysis
of various CE's was first proposed in  \cite{DUKA} in
relation to the calculation of partition functions and Wilson loops of
$QCD_2$ on the sphere, and  developed  further in
\cite{WYNK,GRMA,STWKO}. It appeared to be the only effective method in the
study of the
combinatorics of the so called dually weighted planar graphs (DWG)
\cite{KSW} first introduced in \cite{DIFITZ}.

Although we studied only a particular example of such models
our results suggest  that the universal properties of various correlation
functions observed for our model will hold for many other models defined as
multiple sums over discrete levels (random ensembles on the Young tableaux).

The explicit formulae, exact for finite $N$, which have been obtained
here-above for the density of levels and for the correlation functions,
provide a powerful handle on the study of various universal scaling limits.

It would be nice to apply our methods to
more complicated characteristics of the
model, such as the probability of distribution of the length of various
rows  of the Young tableau, (which in the case of equal $\a$'s
shows, according  to the authors of
\cite{DEIFT}, a remarkable relation to the Painleve II
equation). Unfortunately, for such a quantity,
we have not succeeded to write it in the form of a finite contour integral
representation, similar to that used in our paper for the correlation
functions.


\vskip 14.5pt\hskip -18pt{\Large \bf Acknowledgments}

One of us (V.K.) thanks A. Vershik and L. Pastur for  useful
discussions and valuable bibliographical comments.

\section {Appendix : Density of levels by a direct saddle point calculation}
It is interesting to compare the result obtained in section 5 with a direct
calculation in the
large$N$-limit. For the simplest model in which all the $\a$'s are equal,
the method
is straightforward ; it is based, as usual, on the solution of a
Riemann-Hilbert
problem.

 The probability weight of a given set of random $h$'s, is
proportional by definition to $\displaystyle\a^{\sum h_k}
\Delta(h_1,\cdots,h_N)$ . Is we write it in terms of the density
distribution
\be \rho(x) = \frac{1}{N} \sum_1^N \delta (x-h_k/N),
\ee
this weight is proportional to \\$\displaystyle \exp
N^2\left(\log\a\int dx x\rho(x) + \int dx dy \log\vert x-y\vert
\rho(x)\rho(y)\right).$ The similarity with the unitary ensembles of
random matrices, with their characteristic logarithmic repulsion, is
manifest.  In the large-$N$ limit the weight is maximum for a density
which satisfies the condition \be \log\a + 2 P\int dy
\frac{\rho(y)}{x-y} = 0 \ee.  This condition holds on the support of
the measure, which is yet to be determined. We shall make the ansatz
(proposed in \cite{DUKA} in a similar situation) that $\rho(x)$
remains equal to one for $0<x<b$, is some function of $x$ in the
interval $b<x<a$, vanishes at $x=a$ and remains zero for $x>a$. The
consistency of these conditions will be checked later.  The equation
for $\rho(x)$ on the interval $b,a$ becomes \be \log \sqrt{\a} +
\log\left(\frac{x}{x-b}\right) + P\int_b^{a} dy \frac{\rho(y)}{x-y} =0
\ee We introduce the resolvent \be G(z) = \int_0^{a} dy
\frac{\rho(y)}{z-y}\ee, and \be H(z) = \int_b^{a} dy
\frac{\rho(y)}{z-y}\ee.  The function $H(z)$ is analytic in the
cut-plane from $b$ to $a$. On the cut it satisfies \be \rm{Re} H(x) +
\log \frac{x\sqrt\a}{x-b} = 0.\ee One verifies readily that the
function \be H(z) = \frac{-1}{\pi} \sqrt{(z-a)(z-b)} \int_b^{a}
\frac{dx}{z-x} \frac {1}{\sqrt{(x-b)(a-x)}} \log\left(
\frac{x\sqrt{\a}}{x-b}\right) \ee is the only fuction which satisfies
the analycity requirements and the behaviour at infinity. If we
consider a large circle in the complex u-plane, the behavior at
infinity implies that \be \oint \frac{du}{z-u} \frac
{1}{\sqrt{(u-b)(u-a)}} \log\left( \frac{u\sqrt{\a}}{u-b}\right) =
0. \ee Shrinking now the contour, and collecting the various
singularities, we obtain \be H(z) = -\log \left(
\frac{z\sqrt{\a}}{z-b}\right)- \sqrt{(z-a)(z-b)}\int_0^{b} \frac
{dx}{z-x} \frac{1}{\sqrt{(a-x)(b-x)}}.\ee The integration is then
elementary and we end up with \be G(z) = -\frac{1}{2}\log{\a} -2 \log
\frac{ \sqrt{b(z-a)} +\sqrt{a(z-b)}}{\sqrt{z(a-b)}}.\ee The behavior
at infinity $G(z)\sim \frac{1}{z}$ fixes $a$ and $b$ ; one finds \be
a=\frac{1+\sqrt{\a}}{1-\sqrt{\a}} \rm{and} \ \ b =
\frac{1-\sqrt{\a}}{1+\sqrt{\a}}.\ee One finds then \be G(z)= \log
{\frac{1}{2\a z} \left( (1+\a)z -(1-\a)
-(1-\a)\sqrt{z-a)(z-b)}\right)} ,\ee in agreement with the result of
section 5.  The density of levels is then obtained as \be \rho(x) =
-\frac{1}{\pi} \rm{Im} G(x+i0) , \ee and one verifies our assumptions
: $\rho(x)= 1 \ \rm{for}\ 0<x<b$, vanishes for $x>a$ and in the
interval $(b,a)$ \be \rho(x) = \frac{2}{\pi}
\arctan\sqrt{\frac{b(a-x)}{a(x-b)}}.\ee
This result has been first obtained by A. Vershik \cite {VER}.


\setlength{\baselineskip}{0.666666667\baselineskip}

\end{document}